\documentclass[11pt]{article}
\usepackage{epsfig,amsmath,amssymb,bm,cases,graphicx,psfrag,stfloats,flushend }
\usepackage{epstopdf,subfigure}
\usepackage{color}
\usepackage{url}
\input psfig
\newtheorem{theorem}{Theorem}[section]
\newtheorem{proposition}{Proposition}[section]

\newtheorem{lemma}{Lemma}[section]
\newtheorem{corollary}{Corollary}[section]
\newtheorem{remark}{Remark}[section] 

\def\comment#1{}

\def\P{{\mathbb P}}   
\def\E{{\mathbf E}}
\def\N{{\mathbb N}}
\def\dref#1{(\ref{#1})}
\def\A{{\cal A}}
 
\begin{document}

\begin{center}
{\Large {\bf On the number of active links in random wireless networks}}
\end{center}

\begin{center}
{ \bf Hengameh Keshavarz\footnotemark[1], Ravi R. MAZUMDAR\footnotemark[2], Rahul Roy\footnotemark[3]and Farshid Zoghalchi\footnotemark[4]
}
\end{center}

\noindent \footnotemark[1] Department of Communications Engineering, University of Sistan and Baluchestan, Zahedan, Iran (E-mail: keshavarz@ece.usb.ac.ir)

\noindent \footnotemark[2] Electrical and Computer Engineering, University of Waterloo,
Waterloo, ON  N2L 3G1, Canada (E-mail: mazum@uwaterloo.ca)

\noindent \footnotemark[3] Indian Statistical Institute, New Delhi, India (E-mail: rahul.isid@gmail.com). 

\noindent\footnotemark[4]Department of Mathematics, University of Toronto, Canada (E-mail:.zoghalchi@mail.utoronto.ca)

\begin{center}
\today
\end{center}

\begin{abstract}
This paper presents results on the typical number of simultaneous point-to-point 
transmissions above a minimum rate that can be sustained in a network with $n$ 
transmitter-receiver node pairs when all transmitting nodes can potentially 
interfere with all receivers. In particular we obtain a scaling law when the 
fading gains are independent Rayleigh distributed random variables and the 
transmitters over different realizations are located at the points of a 
stationary Poisson field in the plane.  We show that asymptotically with probability 
approaching 1, the number of simultaneous transmissions (links that can transmit 
at greater than a minimum rate) is of the order of $O(n^{\frac{1}{4}})$. These 
asymptotic results are confirmed from simulations.

\vspace{0.3cm}

\end{abstract}


\section{Introduction}

Consider the situation where there are $n$ transmitter-receiver pairs that are 
randomly distributed over the spatial domain. A transmitter will transmit to its 
designated receiver only if it can deliver a rate greater than a certain 
minimum rate. Otherwise it will choose not to transmit. A natural question to 
ask is what is the number of simultaneous transmitter-links that can exist? Of 
course, in a particular situation, the numbers are dictated by 
the geometry of the transmitter placements. Nevertheless, over all possible 
random configurations we can obtain some insights on the simultaneous number of 
links when the number $n$ is large and the area is finite. This is the question 
that we address in this paper. In particular we show a concentration of 
distribution type of result when the transmitters have a uniform distribution 
over the area. The problem is motivated by networks of base stations that wish 
to communicate to users nearby such that a minimum rate can be guaranteed 
otherwise it does not transmit to reduce the interference in the network. Over 
all realizations, this number is random but there is typicality in behavior. The model can be thought of 
an instance where  cellular towers are located in a given area that transmit to users in their vicinity 
with a given power  and will do so only if they can provide a minimum rate to the user.
If they choose to transmit they will cause interference at receivers of other transmitting towers and the aim is to estimate 
the number of such two-way communications possible as a function of the number of towers distributed 
uniformly over the area over different realizations.

The pioneering work of  Gupta and Kumar \cite{GuptaITT0300} was the first 
concrete approach based on a simple communication model of exclusion called a 
node exclusive model and  exploited a uniform random geometric structure of node 
placement.  In their model, interference only affects the size and geometry of 
exclusion regions. Since then, many researchers have tried to consider more 
realistic situations (i.e. the communication model, link loss model) and  
present tighter throughput bounds. As shown in \cite{Mhatre} the assumption of a 
simple communication model as in \cite{GuptaITT0300} can lead to overly 
optimistic results. Assuming a power-law path-loss model for each 
source-destination pair channel, analysis based on the models considered in 
\cite{GuptaITT0300} and \cite{FranceschettiITT0307}  shows that the per-node 
throughput scales with $\Theta(\frac{1}{\sqrt n})$, where $n$ denotes the total 
number of nodes in the network. Introducing multi-path fading effects, in 
\cite{GowaikarITT0706} the authors assume that the channel gains are drawn 
independently and identically distributed (iid) from a given probability density 
function (pdf). As a particular example, \cite{GowaikarITT0706} shows that the 
throughput scaling law of the Rayleigh fading channel is logarithmic. In 
\cite{EbrahimiITT1007, EbrahimiIT2011}, a rate-constrained single-hop wireless 
network with Rayleigh fading channels is considered.  An upper bound and a lower 
bound of the order $\ln(n)$ on the number of active links supporting a minimum 
rate are obtained. The result is based on a threshold activation policy and the 
idea is to choose a threshold such that the given rate can be achieved. In 
\cite{ToumpisINFOCOM04}, each channel gain is a product of a path-loss term and 
a non-negative random variable modeling multi-path fading and having an 
exponentially-decaying tail. In this case, the achievable per-node throughput 
scales with $\Omega\left(\frac{1}{\sqrt{n(\ln n)^3}}\right)$. In 
\cite{NebatWiOpt06}, the same channel model is considered  and it is shown that 
for a path-loss exponent $\alpha>2$ and any absorption modeled by exponential 
attenuation, a per-node throughput of the order $\Omega(\frac{1}{\sqrt n})$ is 
achievable.

We consider a wireless network of {\em n} transmitter-receiver node pairs where 
any transmitting user can potentially cause interference at a receiver node.The 
aim is to estimate the number of transmitter-receiver pairs that can 
simultaneously exist such that they can transmit at a rate of at least $R_{min}$ 
over random realizations of the transmitter-receiver pairs. We assume that the 
channel gains are due to two components, a fading gain that is Rayleigh 
distributed that we assume is i.i.d. over all channels and a distance based 
attenuation, the path-loss,  that decreases as $d^{-\alpha},\ \alpha > 2$ where 
$d$ is the distance between an interfering transmitter and receiver. The value of 
$\alpha$ is typically 3. We assume that the transmitters are uniformly 
distributed over the domain (made precise later) and the fading gains are 
independent of the location. We show that the number  of simultaneous 
transmissions between transmitters and their receivers is of the order 
$O(n^{\frac{1}{4}})$. Our results differ from earlier ones reported in 
\cite{EbrahimiIT2011} in that they only estimate the number of links that have 
rates above a minimum when all transmitter-receiver pairs are activated. 
Moreover, the geometric aspects were not directly addressed.

The paper is organized as follows: In Section II, the network model is 
introduced. Section III presents the main results where we show that the 
combination of multipath fading and random distance attenuation induces a Pareto 
type of distribution for the interfering gains. In Section IV we conclude with 
some simulation results that confirm the principal result.
We use the following notation: we say $f_n$ is $O(g_n)$ if $\limsup_n 
\frac{f_n}{g_n} < \infty$ and $f(n)\sim g(n)$ means $\displaystyle\lim_{n\to 
\infty}\frac{f(n)}{g(n)}=1$. Similarly for a sequence of random variables 
$\{X_n\}$ and a deterministic function $f_n$ we say that $X_n$ is $O(f_n)$ if 
$\limsup_{n\to\infty} \frac{X_n}{f_n} < \infty\ a.s$. Similarly we say $X_n \sim 
o(f_n)$ a.s.  if $\lim \frac{X_n}{f_n} = 0\ \ a.s.$. We also use the terminology 
$a.a.s$ to refer to a property holding asymptotically almost surely.

\section{Network Model}  \label{model}

\begin{figure}[htbp]
    \begin{center}
    \includegraphics[scale=0.75]{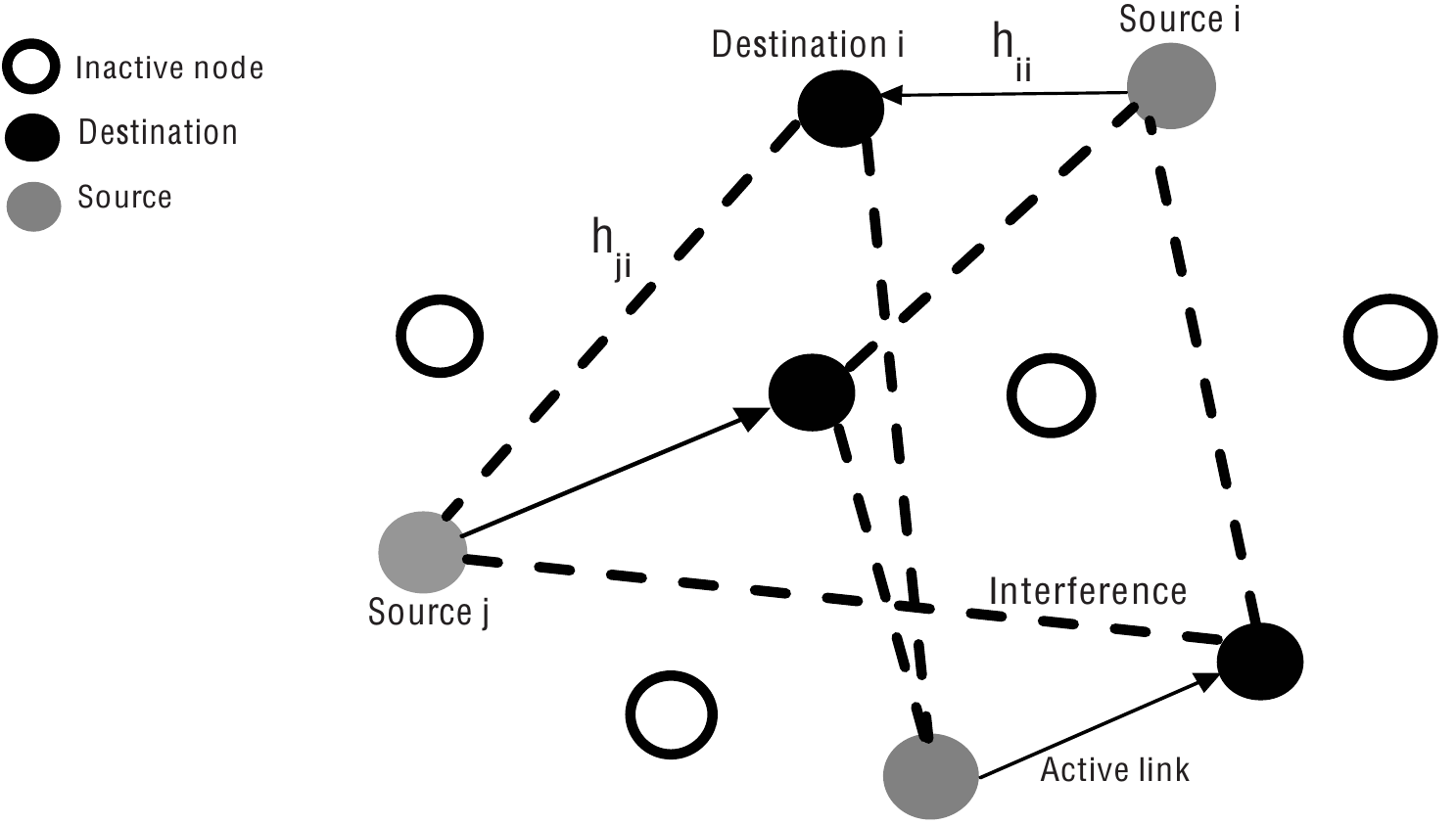}
  \caption{A wireless network with active links $(-)$ and interference channels 
 $(--)$}
   \end{center}
    \label{WNet}
 \end{figure}

Consider a wireless network with $n$  transmitter and receiver pairs as shown in Figure \ref{WNet}. It is 
assumed that a transmitter $i$ transmits to its receiver $i$ through a 
stationary i.i.d fading channel denoted by $h_{ii}$. Transmitting nodes $j\neq 
i$ can interfere with  receiver $i$ and the channel gain between interfering 
transmitters at a receiver $i$ is denoted by $h_{ji}, \ j\neq i$. It is assumed 
that the dominant factor affecting the gain between a transmitter and receiver 
is only due to multipath fading, i.e., distance is ignored, for example at a 
fixed distance from the transmitter. The scenario is one of a receiver being in 
the vicinity of a base station. The rate it receives is only affected by the 
interference from other transmitters that are transmitting to their receivers at 
the same time.

Throughout the paper, by sources and destinations, we mean transmitting nodes 
and receiving nodes respectively. Destinations are conventional receivers 
without multi-user detectors; in other words, no broadcast or multiple-access 
channel is embedded in the network. Nodes transmit signals with maximum power of 
$P$ or remain silent during each time slot.

Let $(t_i, r_i) \in \mathbb R^2 \times \mathbb R^2 $, $i \geq 0$ 
denote the location of the $i$th transmitter-receiver pair. The random model we consider is as follows.

Let $r_1, r_2, \ldots$ be a marked Poisson point process of intensity $n$ on the plane $\mathbb R^2$ with the  receiver $i$ located at $r_i$ having a mark $t_i \in \mathbb R^2$. We assume that $t_i$ depends on $r_i$ however in such a way that  (a) the process  $t_1, t_2, \ldots $ is a Poisson point process on $\mathbb R^2$ of intensity $n$, and, (b) $r_i$ and $t_j$ are independent whenever $i \neq j$. This occurs, for example, when $t_i = r_i + W_i$, where $W_1, W_2, \ldots$ is a sequence of i.i.d. bounded random vectors.

Let $S_1$ be the disc of unit area centered at the origin. From the Palm theory  of Poisson point processes (see Daley and Vere-Jones (1988)\cite[Ch. 12]{Daley}) we know that 
\begin{itemize}
\item[(a)]  the number of transmitters lying in a disc of unit area centered at the location $r_j$ of the $i$th receiver, i.e., $N_j := \#\{t_i: i \neq j, 1 \leq i < \infty\} \cap (r_j + S_1) $,  has a Poisson distribution with mean $n$, and,  
\item[(b)] given $N_j = k$, the $k$ points  $\{t_i: i \neq j, 1 \leq i < \infty\} \cap (r_j + S_1) $ are uniformly distributed in $S_1$. 
\end{itemize}

At the steady state of the system, the signal, $Y_i$, received at receiver $i$, is given by
\begin{eqnarray} \label{net}
 Y_i  = h_{ii}\, X_i + \mathop{\sum_{j\in {\cal A}_i,j\neq i} h_{ji}\, 
X_j} + Z_i 
\end{eqnarray}
where $h_{ii}$ denotes the link fading channel between transmitter $i$ and 
receiver $i$, ${\cal A}_i$ denotes the set of 
active transmitter-receiver pairs in the unit area neighbourhood of the $i$th receiver at $r_i$ and $Z_i \sim {\cal CN}(0,\sigma^2)$ 
represents background noise at node $i$ during a time-slot. Note $X_i = P$ if $i$ is 
transmitting and 0 otherwise.

We assume that the channel activation from slot to slot is independent and each 
node uses a threshold based strategy  for activation, .i.e.,  $X_i = P$ if 
$|h_{ii}|^2 > h_0$ where $h_0$ is a threshold. This is referred to as a TBLAS  
(Threshold Based Activation Strategy) in \cite{EbrahimiIT2011}.  Only 
nodes that can sustain a given rate of transmission are activated and thus such 
a strategy is not fully decentralized. However, this allows us to obtain an 
estimate of the largest number of concurrent activated links.  Let  $\N_n= \{ j: 1 
\leq j \leq N_i\}$ denote the set of possible links in the unit area region $r_i+S_i$ centered at $r_i$, where $\N_i$ is the Poisson number 
of transmitters in this region, as described earlier.

The achievable rate bits of link $i$ can be thus be written as
\begin{eqnarray}  \label{rate}
R_i  &\leq& B\ln \left(1+ \frac{P \left|h_{ii}\right|^2}{\sigma^2 + 
\displaystyle\mathop{\sum_{j\in {\cal A}_i, j\neq i} P \left|h_{ji}\right|^2}} 
\right)   
\end{eqnarray}
where $B$ is the spectrum bandwidth.

The above is an equality if the noise and channel gains are gaussian which is 
the case here. For convenience we take $B=1$ throughout the paper.

In the remainder of the paper our goal is to estimate $m= \#
|{\cal A}_i|$, the cardinality of ${\cal A}_i$  with the maximum number of simultaneous 
transmitting links that can exist at a time when we require that all transmitters in ${\cal A}$ 
must be able to transmit at a rate greater than $R_{min}$. Note  ${\cal A}_i \subset 
\N_n$ and hence $m$ depends on $n$ , i.e., $m=m(n)$, is random and depends on 
the channel gain realization. We show that when $n$ is large the distribution of 
$m(n)$ sharply concentrates around a given value modulo constants.

Let us denote by $\gamma_i(n)$:
\begin{equation}
\gamma_i(n)= B\ln \left(1+ \frac{P \left|h_{ii}\right|^2}{\sigma^2 + \displaystyle\mathop{\sum_{j\in {\cal A}_i, j\neq i} P \left|h_{ji}\right|^2}} \right)   
\end{equation}

\section{Main results}

In delay-sensitive applications, each active link needs to support a minimum rate. Due to limited transmitted power and interference from other active source-destination pairs, it is not always possible for all nodes to keep this minimum rate. Hence, only nodes with good channel conditions should be active while others remain silent during each time slot. Consider the received signal model given by \dref{net}. 

Let $\A_m$ denote the set  of active transmitters. Define the stochastic rate $\gamma_{i,m}$ of link $i$ as

\begin{eqnarray} \label{x}
\gamma_{i,m}  &\stackrel{\Delta}{=}&  \ln\left(1+ \frac{P \left|h_{ii} \right|^2 {\bf 1}_{[|h_{ii}|^2 > h_0]}}{\sigma^2 + \displaystyle\mathop{\sum_{j\in \A_m, j\neq i} P \left|h_{ji}\right|^2 {\bf 1}_{[|h_{jj}|^2 \geq h_0]}}}  \right)  
\end{eqnarray}

Then the maximum number of active links supporting the minimum rate is given by the following optimization problem.

\begin{numcases}{}
M_n=\max |\A_m|, \ m\leq N_n     \label{mm} \\
\gamma_{i,m} \geq R_{\min}, \quad i\in \A_m \label{Ri},
\end{numcases}
Clearly, with fixed $P$ and $R_{\min}$, the maximum number of active links is a random variable which depends on the Poisson point process, the channel gains $\left|h_{ij}\right|^2$; $i,j=1,\ldots, n$ and the interference caused by nodes transmitting at the same time.

In wide area networks attenuation due to path-loss plays a significant role in determining the quality of the link, i.e., the rate at which a transmitter-receiver pair can communicate. In our model the channel gain between a transmitter-receiver pair $i$ denoted by $h_{ii}$ is due to fading only with Gaussian channel conditions.  This leads to channel gains between T-R pairs to be Rayleigh distributed. In the sequel we use $h_{ij}$ to denote $h_{ij}^2,\ \ i,j \in \{1,2,\ldots,n\}$. Without loss of generality we assume that $h_{ii}$ has an exponential distribution with mean 1.

Accounting for both fading and path-loss between a transmitter $i$ and receiver $j$ for $i\neq j$  is characterized by a channel gain of the form:
\begin{equation}
\label{def:h_{ij}}
h_{ij} = g_{ij} D^{-\alpha}_{ij} 1_{\{D_{ij} \leq 1\}},\ \ \alpha \geq 2
\end{equation}
where $g_{ij}$ is the fading gain which we assume to be  an exponential random variable with mean 1 (Chi-squared with two degrees of freedom and mean 1),   $D_{ij}$ is the distance from transmitter $i$ to receiver $j$ and $\alpha$ is the path loss exponent which is typically 3. 
In the expression (\ref{def:h_{ij}}), the indicator function guarantees that there is no effect on a receiver from a transmitter at a distance 1 or more from it. From the scaling properties of the Poisson process we may deduce that this restriction is minor. Indeed, the transformation $x \mapsto ax$ applied to a Poisson process of intensity $n$, keeps its Poissonness intact and just changes the intensity to $n/ (a)^2$. Thus a bound on the radius of influence of a receiver may be adjusted with a corresponding change in intensity of the Poisson process.
Over different  realizations the distances $D_{ij}$ are assumed to be random and denote the distances from transmitter $i$ to receiver $j$ noting that the transmitters (and receivers) form a stationary Poisson field with intensity $n$. A similar model has also been considered Baccelli and Singh \cite{Baccelli} in the context of spatial random access schemes.

In the following we denote by $g$ the density $g_{ij}$ of the exponential random variable with mean 1 representing the fading gain between  a transmitter $i$ and receiver $j$   ( $i \neq j$), and by $D$ the random variable with the same distribution as each of the i.i.d.  distance random variables of $D_{ij}$ ( $i \neq j$) from transmitter $i$ to receiver $j$ whose distribution is obtained from the spatial distribution of the transmitter-receiver pairs. We assume that $D_{ij}$ is independent of the random variable whose density is $g_{ij}$.

\begin{lemma}
\label{heavygain}
Fix $b > 0$. For any receiver $j$ and transmitter $i$ and $h_{ij}$ as above,  we have
\begin{equation}\label{heavy}
c_1z^{-\beta} \geq \P(h_{ij}> z | t_i \in r_j + S_1) \geq c_2 z^{-\beta}\text{ for all } z\geq b,
\end{equation}
where $\beta= \frac{2}{\alpha}$ and $c_1= c_1(\alpha) > 0,\ c_2= c_2(\alpha, b)>0$ are constants that depend on $\alpha$ but are bounded, i.e. $h_{ij}$ is  heavy-tailed.
\end{lemma}

\noindent{\bf Proof:}
Without loss of generality, suppose the receiver $i$ is located at the origin, so that there are $N_i$ transmitters in the region $S_1$ which have an effect on the receiver $i$, where $N_i$ is a Poisson random variable of mean $n$ as discussed in (b) of the description of the model.

Given $t_i \in S_1$, we know that the transmiter $i$ is uniformly located in the ball $S_1$, so that  
$\P(D_{ij} \leq u) = u^2= \frac{\pi s^2}{\pi}$ for $0 \leq  u\leq 1$.with the density of $D_{ij}$ being  $p_d(u)=2u1_{\{0 \leq  u\leq 1\}}$

Therefore,
\begin{eqnarray*}
\P(h_{ij} > z| t_i \in r_j + S_1) & = & \P(g_{ij}D_{ij}^{-\alpha} > z| t_i \in r_j + S_1) = \P(g_{ij}> zD_{ij}^{\alpha}| t_i \in r_j + S_1)\\
& = & \int_0^1 \P(g_{ij}>zu^{\alpha}|t_i \in r_j + S_1)p_d(u)du\\
&=& 2\int_0^1 ue^{-zu^{\alpha}}du
\end{eqnarray*}
where we have used fact that $\P(g_{ij}> x) = e^{-x}$ for $x > 0$.

Now,
\begin{eqnarray*}
 \int_0^1u e^{-zu^{\alpha}}du & = & z^{-\frac{2}{\alpha}} \int_0^z v^{\frac{2}{\alpha}-1} e^{-v}dv\\
 & = & \Gamma(\frac{2}{\alpha}) P(\frac{2}{\alpha}, z) z^{-\frac{2}{\alpha}}
 \end{eqnarray*}
 where$\Gamma(a)$ is the Gamma function and $P(a,x)$ is the Incomplete Gamma function for $a,x>0$. Note that by definition $P(a,0) < P(a,x) \leq P(a,\infty)=1$. This completes the proof.

Now, from the fact the the points are independent, the above shows that the channel gains from the interferers are i.i.d. distributed as (\ref{heavygain}). From (\ref{heavygain}) we see that the distribution of the channel gains is a generalized Pareto distribution. Thus the classical Strong Law of Large Numbers (SLLN) does not apply due to the infinite mean since $\alpha \geq 2$ and typically is 3 for the far field model in wireless communications. However, a suitable normalization of the partial sums of heavy-tailed random variables can be associated with a SLLN due to Marcinkiewicz and Zygmund  \cite[Theorem 2.1.5]{EKM97} that we state below.

\begin{theorem}
\label{heavySLLN}
 Let $p\in (0,2)$ and $S_n= \sum_{i=1}^n X_i$ where $\{X_i\}$ are i.i.d. Then the following SLLN holds:
\begin{equation}
\label{SLLN}
n^{-\frac{1}{p}} (S_n-an) \rightarrow 0,\ \ a.s.
\end{equation}
 for some real constant $a$ if and only if $\E|X_0|^p < \infty$. 

If $\{X_i\}$ satisfy (\ref{SLLN}) and $p<1$  we can choose $a=0$ while if $p\in (1,2)$ then $a= \E[X_1]$

\end{theorem}

In our context we need a slight generalization of the above result. First we note that
\begin{lemma}
\label{condSLLN}
Let $X$ be a non-negative  r.v whose tail distribution is given by (\ref{heavy}) with $\alpha \geq 2$. Then for any $0< p < \frac{2}{\alpha}$,  the r.v. $X^p$ is integrable, i.e.
\begin{equation}
\E[|X|^p] < \infty
\end{equation}
\end{lemma}

\noindent{\bf Proof:}

The proof readily follows from the fact that for $0< p <\frac{2}{\alpha} < 1$
\begin{eqnarray*}
\P(X^p > z) & = & \P(X \geq z^{\frac{1}{p}})\\
& \geq  & c_2 z^{-\frac{2}{\alpha p}} = c_2 z^{-(1+\delta)}, \ \mathrm {for\  some\ \delta > 0}.
\end{eqnarray*}

Hence  $X^p$ has a finite mean.

\begin{corollary}
Let $X_1, X_2, \ldots$ be i.i.d. non-negative random variables, each having a probability density function whose tail distribution is given by (\ref{heavy}) with $\alpha > 2$ and $S_n$ as in Theorem \ref{heavySLLN}. Let $N_n$ be a Poisson random variable with mean $n$, independent of the random variables $X_1, X_2, \ldots$. Then we have
\begin{equation}
\label{Poi_SLLN}
N_n^{-\frac{1}{p}} S_{N_n} \rightarrow 0,\ \ a.s., \ \ \forall \ 0< p<\frac{2}{\alpha} < 1
\end{equation}
\end{corollary}
\noindent{\bf Proof:}
Note that
$$
\frac{S_{N_n}}{N_n^{1/p}}  = \frac{S_n}{n^{1/p}} - ( \frac{S_n}{n^{1/p}}  - \frac{S_{N_n}}{N_n^{1/p}}).
$$
Now writing $\frac{S_n}{n^{1/p}}  - \frac{S_{N_n}}{N_n^{1/p}}$ as
$( \frac{S_n}{n^{1/p}}  - \frac{S_{N_n}}{N_n^{1/p}})1_{\{|N_n - n|
\leq \sqrt{n}\ln n\} }+ ( \frac{S_n}{n^{1/p}}  - \frac{S_{N_n}}{N_n^{1/p}})1_{\{|N_n - n| > \sqrt{n}\ln n\}}$ we note that\\
(i) by Theorem \ref{heavySLLN},
$$
( \frac{S_n}{n^{1/p}}  - \frac{S_{N_n}}{N_n^{1/p}})1_{\{|N_n - n| \leq \sqrt{n}\ln n\} }\leq
2\frac{\sum_{j = n-\sqrt{n}\ln n}^{n+\sqrt{n}\ln n} X_i}{n^{1/p}}
\to 0 \text{ almost surely as } n \to \infty ,
$$
(ii) by Chebychev's inequality
$$
P\{|N_n - n| > \sqrt{n}\ln n\}  \leq \frac{\text{Var}(N_n)}{n(\ln n)^2} = \frac{1}{ (\ln n)^2}\to 0
\text{ as } n \to \infty
$$
thus an application of  Slutsky's theorem (see Grimmett and Stirzaker\cite[ p 318]{Grimmett}) completes the proof of the corollary.

We now state the main result of this paper.

\begin{proposition}(Main Result)
\label{Heavythm}
Consider a dipole random SINR graph whose channel gains between transmitters $i$ and $j$ with $i\neq j$  are i.i.d. and whose tail distribution is given by (\ref{heavy}) and the direct channel gain $h_{ii}$ is exp(1) distributed arising from a Rayleigh fading model. Let $\eta_n$ denote the number of simultaneous transmitter-receiver pairs that can transmit at a rate of at least $R_{min}$. Then,
\begin{equation}
\eta_n \sim n^{\frac{1}{4}}\ \ a.a.s. 
\end{equation}
\end{proposition}

We prove the result through showing several intermediate results.

Let $h_0>0$ be a threshold and define the Bernoulli random variables:
\begin{align}
\xi_{j} = \left\{ {\begin{array}{*{20}c}
   {1} & {{\rm{if~~}}h_{ii} > h_0 }  \\
   {0} & {{\rm{if~~}}h_{ii} < h_0 }  \\
\end{array}} \right.
\end{align}
and let: $M_n= \sum_{i=1}^{N_i} \xi_i$ denote the number of {\em good} or potentially active channels. Let $p_0= \P(\xi_i =1)= \P(h_{ii} > h_0)$ and choose $h_0=  \gamma \ln n$  for  $0<\gamma< 1$. Then we can show the following result: 

\begin{lemma}
\normalfont Let $M_n= \sum_{i=1}^{N_i} \xi_i$ where $\{\xi_i\}$ are i.i.d. $\{0,1\}$, random variables with ${\bf E}[X_i]=p_0= e^{-h_0}$ where $h_0=\gamma \ln n$  for  $0 <\gamma < 1/2$.
Then  as $ n\to \infty$
\begin{equation}
\label{goodchnum}
\P(M_n = O(n^{1-\gamma})) \to  1.
\end{equation}

\end{lemma}

\noindent{\bf Proof:}
\begin{eqnarray} \label{bernoulli}
\xi_i=\left\{
\begin{array}{l}
1, \quad \ \mbox{with probability}\ p_0 \\
0, \quad \  \mbox{with probability}\ 1-p_0\end{array}
\right.
\end{eqnarray}
for $i=1,2,\ldots,n$. Then, the number of ``good" links has the same distribution as $M_n=\sum_{i=1}^{N_n} \xi_i$, which satisfies the Binomial distribution $B(N_n,p_0)$. 
\begin{eqnarray}  \label{P_0}
p_0 = \P(h_{ii} >h_0)  &=& \exp\left(-h_0\right)  \nonumber\\
& = & \frac{1}{n^{\gamma}}
\end{eqnarray}
Hence $np_0 = n^{1-\gamma}$.

Now from the fact that $\xi \in \{0,1\}$ , using Hoeffding's inequality, see \cite[Example 8, p 477]{Grimmett} :
\begin{eqnarray*}
\P( |M_n-N_np_0|> \varepsilon_n ) & = &\sum_{k=0}^\infty \P( |M_n-kp_0|> \varepsilon_n )\frac{e^{-n}{n^k}}{k!}\\
&\leq &\sum_{k=0}^\infty  \exp(- \varepsilon_n^2/ (2k(1-n^{-\gamma})))\frac{e^{-n}{n^k}}{k!}.
\end{eqnarray*}

Now let $\varepsilon_n =n^a$ for some $a \in (1/2, 1-\gamma)$ and, for a given $\eta > 0$,  let $n_0$ be large such that for all $n \geq n_0$
(i)  $\exp(- \varepsilon_n^2/ (2n\log n(1-n^{-\gamma}))) < \eta$ and (ii) $ \sum_{k=n\log n}^\infty  \frac{e^{-n}{n^k}}{k!} < \eta$.
Thus, for $n \geq n_0$,
$$
\sum_{k=0}^{n\log n}  \exp(- \varepsilon_n^2/ (2k(1-n^{-\gamma})))\frac{e^{-n}{n^k}}{k!} +  \sum_{k=n\log n}^\infty  \frac{e^{-n}{n^k}}{k!} \leq 
\eta \sum_0^{\infty}\frac{e^{-n}{n^k}}{k!}  + \eta =2\eta
$$
where we used the fact that $ \exp(- \varepsilon_n^2/ (2k(1-n^{-\gamma}))) \leq \exp(-n^{2a}/(2n\log n (1-n^{-\gamma})))$ for $k < n\log n$.
Therefore we have $\P(|M_n-np_0|> \varepsilon_n)\leq 2 \eta $ for $n \geq n_0$. 
Since $\eta > 0$ is arbitrarily small, we have $\P(|M_n - N_np_0|> \varepsilon_n)\  \rightarrow 0  \text{ as } n\rightarrow \infty$. The result is established by Slutksy's theorem and on noting that  as $n \to \infty$, by the strong law of large numbers, $N_n/n \to 1$ and $\frac{\varepsilon_n}{n^{1-\gamma}} \rightarrow 0$.

Next we show that the minimum rate constraint is satisfied for at least $n^\delta$ transmitter-receiver pairs in $\A$ for any $0 < \delta< \gamma$. 

\begin{lemma}
Consider a dipole SINR random graph with $n$ transmitter-receiver pairs. Suppose the channel gains are direct channel gains $h_{ii}$ are $\exp(1)$ distributed and the cross transmitter-receiver channel gains denoted by $h_{ij}, i\neq j$ are i.i.d with distribution given by (\ref{heavy}). Let $\A_m\subset \N_n$ denote the set of $m$ active transmitter-receiver pairs. Then, asymptotically almost surely,   every set $\A_m$ of cardinality $m=n^{\delta}$  with $0 < \delta< \gamma<\frac{1}{2}$ can support a mimimum rate $R_{min}$.
\end{lemma}

Define the set:
\begin{equation}
\label{SLLNregion}
{\cal U}_{\varepsilon, m} = \{ \omega: \frac{1}{m^{\frac{1}{p}}} \sum_{j\in \A_m, j\neq i} h_{ji}{\bf 1}_{[h_{jj} > h_0]} \leq \varepsilon\}
\end{equation}

Clearly for $p < \frac{2}{\alpha}$ by Theorem \ref{heavySLLN}  $\P({\cal U}_{\varepsilon,m}) \rightarrow 1\ as\ m\to \infty$. However we need the following estimate of  probability of the complement of  ${\cal U}_{\varepsilon,m}$. First note that the r.v.'s $h_{ji}{\bf 1}_{[h_{jj}>h_0]}$ are i.i.d. for $j\neq i$ and moreover 
\begin{equation}
\P(h_{ji}{\bf 1}_{[h_{jj} > h_0]} > z) = \P(h_{ji} >z)\P(h_{jj} > h_0) \sim \frac{c_1}{z ^{\frac{2}{\alpha} }n^{\gamma}}
\end{equation}
by independence of $h_{ji}$ and $h_{jj}$ for $j\neq i$. This shows that the random variables $h_{ji}{\bf 1}_{[h_{jj} > h_0]}$ are also heavy tailed with the same exponent $-\frac{2}{\alpha}$.

Now we use the {\em principle of the single large jump} for heavy tailed random variables \cite[Chapter 3]{Foss}

\begin{theorem}
\label{convtails}
Let $\{X_i\}_{i=1}^n$ be a collection of n i.i.d. sub-exponential distributions with common distribution $F(x)$. Then:
\begin{equation}
\P( X_1 + X_2+\cdots +X_n > x) \sim \P (\max_{1\leq i \leq n} X_i > x) \sim 1-F(x)^n \sim n (1-F(x))\ as\ x\to\ \infty
\end{equation}
\end{theorem}

Applying Theorem \ref{convtails} to $\sum_{j\in \A_m, j\neq i} h_{ji} {\bf 1}_{[h_{jj} > h_0]}$ for every fixed $\varepsilon >0,\ \ {\mathrm and} \  m(n)\to \infty\ as\ n\to \infty$ we obtain:
\begin{eqnarray}
\P({{\cal U}^c}_{\varepsilon, m})=  \P(\Omega / {\cal U}_{\varepsilon, m} )& \sim& m \P(h_{ji} {\bf 1}_{[h_{jj} > h_0]} > m^{\frac{1}{p}} \varepsilon) \nonumber\\
& \sim &m \frac{c_1}{(m^{\frac{1}{p}} \varepsilon)^{\frac{2}{\alpha}}n^{\gamma}}\nonumber\\
\label{LDtail}
&\sim& c_1 m^{1-\frac{2}{p\alpha}}{\varepsilon}^{-\frac{2}{\alpha}} n^{-\gamma} \rightarrow 0\ as \ n\to \infty\\
\end{eqnarray}
Note that $p\alpha < 2$ and hence $1-\frac{2}{p\alpha} < 0$

Let us show that if $i \in \A_m$ when $m\sim n^{\delta}, \delta < 1$ then the minimum rate constraint is met when $\omega \in {\cal U}_{\varepsilon, n}$.

Let $X_{i,m}$  be the (random) rate as defined before in (\ref{x}). Now, choose $\varepsilon = \gamma e^{-R_{min}} n^{-\frac{\delta}{p}} \ln n $.  Then,  since $n^{\frac{\delta}{p}} \varepsilon  \to \infty$ the conditions of Theorem \ref{condSLLN} and (\ref{LDtail}) are met. Without loss of generality let us take  the transmit power $P=1$

\begin{eqnarray*} \label{hx}
X_{i,m}{\bf 1}_{[{\cal U}_{\varepsilon, n}]}  &\stackrel{\Delta}{=}&  \ln\left(1+ \frac{ h_{ii}  {\bf 1}_{[h_{ii} > h_0]}}{\sigma^2 + \displaystyle\mathop{\sum_{j\in \A_n, j\neq i}  h_{ji} {\bf 1}_{[h_{jj} \geq h_0]}}} \right) {\bf 1}_{[{\cal U}_{\varepsilon, n}]}  \\
& \geq & \ln\left(1 + \frac{h_0}{\sigma^2+ (m-1)^{\frac{1}{p}} \varepsilon}\right){\bf 1}_{[{\cal U}_{\varepsilon, n}]} \\
& \sim & \ln \left( 1 + \frac{\gamma \ln n}{\gamma e^{-R_{min}} \ln n}\right)\ -a.s.\ \ n\to\ \infty \\
& \sim & \ln (1 + e^{R_{min}}) \geq R_{min}\ a.s.\ \ \ n\to\ \infty
\end{eqnarray*}
since  ${\bf 1}_{[{\cal U}_{\varepsilon, n}]} \to 1\ \ a.s. \ n\to \infty$ by the SLLN given in Theorem \ref{condSLLN}.

Let us now show that indeed $n^{\delta}$ transmitter -receiver pairs can simultaneously transmit above the rate $R_{min}$ provided $\delta \leq \frac{\gamma}{2}$ thus completing the proof of the main result.

First of all, in light of the above result, it follows that:
$$\{\omega: X_{i,m} < R_{min}\} \subset {\cal U}_{\varepsilon,m}^c$$
where $A^c$ denotes $\Omega / A$.

Therefore noting:
$$
\P\left(X_{i,m} < R_{min}\right)  \leq  \P\left( {\cal U}_{\varepsilon,m}^c \right)
$$
from the union bound with $\varepsilon=\gamma e^{-R_{min}} n^{-\frac{\delta}{p}} \ln n$
\begin{eqnarray}
\P\left(\bigcup_{i\in \A_m} \{X_{i,m} < R_{min}\}\right) & \leq & \sum_{i\in \A_m} \P (X_{i,m} < R_{min})\nonumber \\
\label{a}
&  \leq & m \P\left( {\cal U}_{\varepsilon,m}^c\right) \\
\label{b}
& \leq & c_1 m^2(m^{\frac{1}{p}}\varepsilon)^{-\frac{2}{\alpha}} n^{-\gamma} \\
\label{ca}
& \leq & const. n^{2\delta -\gamma} \frac{1}{(\ln n)^{\frac{2}{\alpha}}} \rightarrow 0
\end{eqnarray}
where (\ref{a}) follows from the union bound and fact that the $X_{i,m}'s$ are identically distributed, (\ref{b}) follows from (\ref{LDtail}) and (\ref{ca}) follows by our choice of $\varepsilon$.

Since $n^{\delta}$ denotes the cardinality of the set of ``good"  transmitters and it implies that $\delta \leq \frac{\gamma}{2}$, $\gamma< \frac{1}{2}$    and therefore $\delta < \frac{1}{4}$ and we can make it as close to $\frac{1}{4}$ as needed.

The proof of the upper-bound can be obtained by noting that when the direct fading gains are Rayleigh,  $\max_{1\leq i\leq n} h_{ii} \sim \ln n$. Therefore if $\gamma> 1$ the cardinality of the "good" set of probable links goes to zero.  From Lemma \ref{goodchnum}, $\gamma  < \frac{1}{2}$, Then,  it can be seen that our estimate of $n^{\delta}, \delta \leq \frac{\gamma}{2}$ with $0 < \gamma <\frac{1}{2}$ is maximal in that if the cardinality is higher then asymptotically the rate constraint cannot be met. This completes the proof of the result.

\begin{remark}
The results  rely on the independence hypothesis of the channel gains. If we consider a simplified model with i.i.d. Rayleigh fading ignoring the geometric aspects of the problem (i.e. ignoring path loss) the it can be shown the typical number of rate constrained links is $\sim\ (\log n)^2$ which is much lower than the reported result. Thus spatial aspects help improve the total communication rates due to path loss effects making interference from more distant transmitters be negligible.  This scaling law gives an idea of typical behavior over many realizations of the wireless system due to placement of $n$ transmitters in a bounded region.\end{remark}

\section{Simulation Results}\label{Paretosimu}

 We simulated a dipole random model presented in section \ref{model} and assumed that the T-R channel, i.e. the $h_{ii}$ gains are i.i.d. $\exp(1)$ and  the interfering channel gains, $h_{ij}, \ i\neq j$ are i.i.d, Pareto with $\alpha=3$.  The maximum transmitted power was taken as $P=0.032$ watt (i.e. 15 dBm which is typical power for WiFi). The spectrum bandwidth is $B=22$ MHz (typical for WiFi) and the background noise variance is $\sigma^2=0.01$.  Numerical results on each figure were generated by Monte-Carlo simulations.

\begin{figure}[ht]
	\centering
	{\subfigure[Active links vs. total for $R_{min}=100kbps$]{
  {\psfig{file=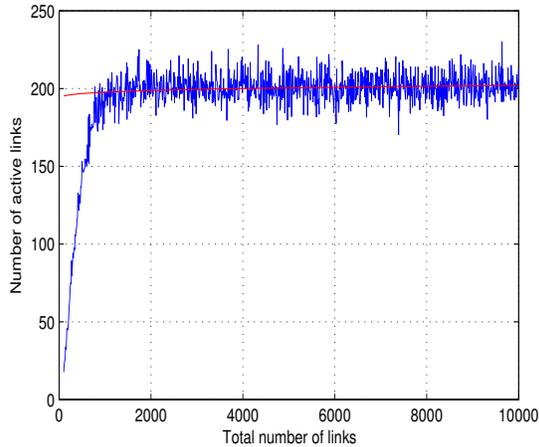,width=3.1in,height=2.5in}
	    \label{Rmin100}}}
	\subfigure[Active links vs. total for $R_{min}=150kbps$]{
	    \psfig{file=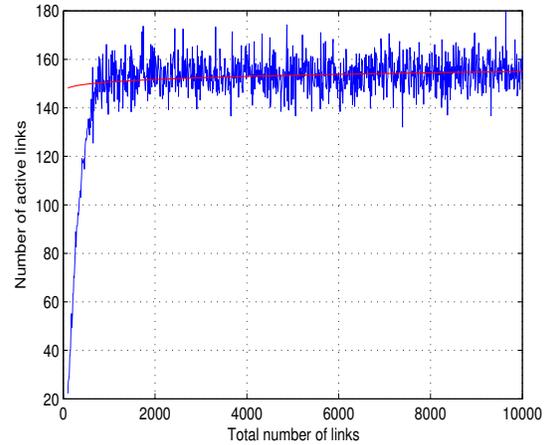,width=3.1in,height=2.5in}
	    \label{Rmin150}}
	\begin{center}{\caption{Number of active links vs. total number for different minimum rates}}\end{center}
	\label{activelinks}}
	\end{figure}

Figure \ref{Rmin100} shows the number of links supporting  a minimum rate of $100$ Kbps versus the total number of possible T-R pairs. Both simulation results (in blue) and the theoretical estimate (in red) shifted by an additive constant given by Proposition \ref{Heavythm} are indicated on this figure. It can clearly be seen that there is a constant gap between the numerical and theoretical results as seen from the simulation results that are centered around the line $C_1+   n^{\frac{1}{4}}$ where $C_1$ is a constant.

Likewise, Figure \ref{Rmin150}  shows the number of links supporting  a minimum rate versus the total number of users for $R_{min}=150$ Kbps .Once again we see that the asymptotic $C_1(R_{min}) +  n^{\frac{1}{4}}$ provides a very good estimate of the number of simultaneous T-R pairs when there are more than 100 T-R pairs. For the case of $R_{min}= 100Kbps$  the additive constant is $C_1= 192$ while for the case $R_{min}=150Kbps$, the constant is given by $C_1= 145$. It is not difficult to see that the constant $C_1$ should be inversely proportional to $R_{min}$.

\section{Acknowledgment}

This work was supported by Natural Sciences and Engineering Research Council (NSERC) of Canada. RM would like to acknowledge the support and hospitality of LINCS (Laboratory of Information, Networks and Communication Sciences) and INRIA-ENS, Paris.


\begin{thebibliography}{9}

\bibitem{GuptaITT0300}
P. Gupta and P. R. Kumar, ``The capacity of wireless networks", 
{\em IEEE Trans. Information Theory}, vol. 46, no. 2, pp. 388-404, March 2000.

\bibitem{FranceschettiITT0307}
M. Franceschetti, O. Dousse, D. N. C. Tse, and P. Thiran, ``Closing the gap in 
the capacity of wireless networks via percolation theory", 
{\em IEEE Trans. Information Theory}, vol. 53, no. 3, pp. 1009-1018, March 2007.

\bibitem{GowaikarITT0706}
R. Gowaikar, B. Hochwald, and B. Hassibi, ``Communication over a wireless 
network with random connections", 
{\em IEEE Trans. Information Theory}, vol. 52, no. 7, pp. 2857-2871, July 2006.

\bibitem{ToumpisINFOCOM04}
S. Toumpis and A. J. Goldsmith, ``Large wireless networks under fading, 
mobility, and delay constraints", 
{\em IEEE Conf. Computer Communications (INFOCOM)}, pp. 609-619, Hong Kong, 
China, March 2004.

\bibitem{NebatWiOpt06}
Y. Nebat, ``A lower bound for the achievable throughput in large random wireless 
networks under fixed multipath fading", 
{\em Intern. Symp. Modeling and Optimization in Mobile, Ad Hoc and Wireless 
Networks (WiOpt)}, pp. 1-10, Boston, USA, April 2006.


\bibitem{Daley}
D.J. Daley and D. Vere-Jones, {\em An introduction to the theory of point processes}, Springer Series 
in Statistics, Springer-Verlag, New York,1988.

\bibitem{Baccelli}
F.  Baccelli and C. Singh,  ``Adaptive Spatial Aloha, Fairness and Stochastic Geometry", CoRR, 2013. 
Available from: http://arxiv.org/abs/1303.1354.


\bibitem{Gut05}
A. Gut, {\em Probability: A Graduate Course}, Springer, 2005.

\bibitem{Mhatre}
V. P. Mhatre, C. P. Rosenberg, and R. R. Mazumdar,`` On the capacity of ad hoc 
networks under random packet losses", {\em IEEE Trans. on Information Theory}, 
Vol 55 (6), 2009, pp 2494-2498.


\bibitem{Grimmett}
G.R. Grimmett and D. R. Strizaker, {\em Probability and Random Processes}, 3rd. 
Ed, Oxford Science Publ.,  2001.

\bibitem{EbrahimiITT1007}
M. Ebrahimi, M. A. Maddah-Ali, and A. K. Khandani, "Throughput scaling laws for 
wireless networks with fading channels," {\em IEEE Trans. Information Theory}, 
vol. 53, no. 11, pp. 4250 – 4254, November 2007.

\bibitem{EbrahimiIT2011}
M. Ebrahimi and A. K. Khandani, "Rate-constrained wireless networks with fading 
channels: Interference-limited and noise-limited regimes," 
{\em IEEE Trans. Information Theory}, Vol 57 (12), 2011, pp. 7714-7731.

\bibitem{HaenggiITT1005}
M. Haenggi, ``On distances in uniformly random networks", 
{\em IEEE Trans. Information Theory}, vol. 51, no. 10, pp. 3584-3586, October 
2005.

\bibitem{Graham94}
R. L. Graham, D. E. Knuth, and O. Patashnik, {\em Concrete Mathematics: A 
Foundation for Computer Science}, Second Edition, Reading, Massachusetts: 
Addison-Wesley, 1994.

\bibitem{Rohatgi76}
V. K. Rohatgi, {\em An Introduction to Probability Theory and Mathematical 
Statistics}, John Wiley \& Sons, Inc., New York, 1976.

\bibitem{FranceschettiAllerton07}
M. Franceschetti, M. D. Migliore, and P. Minero, 
``The capacity of wireless networks: information-theoretic and physical limits", 
IEEE Trans. on Information Theory, 55(8), pp. 3413-3424, August 2009

\bibitem{EKM97}
P. Embrechts, C. Kluppelberg, and T. Mikosch, {\em Modelling Extremal Events for 
Insurance and Risk}, Springer-Verlag, Berlin, 1997.

\bibitem{Foss}
S. Foss, D. Korshunov, and S. Zachary;
{\em An introduction to heavy-tailed and subexponential distributions.} Springer 
Series in Operations Research and Financial Engineering. Springer, New York, 
2011. 


\end{thebibliography}
\end{document}